\documentstyle[epsfig]{aipproc}

\def\fnum@figure{}
\begin{document}

{\hspace*{\fill} 'T is strange, but true; for truth is always strange.\\
\hspace*{\fill} {\it Lord Byron}}
\title{Suppression and restoration of superconductivity in PrBa$_{2}$Cu$_{3}$O$%
_{7}$.}
\author{I.I. Mazin}
\address{Code 6391, Naval Research Laboratory,
4555 Overlook Ave. N.W., Washington, DC 20375}
\maketitle
\vskip 5mm
One of the most exciting cases of superconductivity suppression in high-$%
T_{c}$ cuprates is that of $RE_{1-x}$Pr$_{x}$Ba$_{2}$Cu$_{3}$O$_{7},$ where $%
RE$ stands for a rare earth (see Refs.\cite{rad,PM} for reviews). Even more
exciting are indications that conductivity and superconductivity can
be restored in stoichiometric PrBa$_{2}$Cu$_{3}$O$_{7}$\cite{newT}.
This is such an unexpected result\cite{lee} that it is still not generally
accepted and further experimental confirmation is required. Nevertheless,
this fact was reported by several independent groups, and it is time now to
understand the theoretical consequences of this finding. The most important
message, if this finding is true, is that at $x=1$, and, presumably, at
intermediate $x$'s, there are free carriers in $RE_{1-x}$Pr$_{x}$Ba$_{2}$Cu$%
_{3}$O$_{7}$, and the suppression of metallic conductivity at sufficiently
large $x$ must be due to localization of those carriers. In this paper I
review the principal experimental findings about superconductivity,
suppression thereof, and related properties of $RE_{1-x}$Pr$_{x}$Ba$_{2}$Cu$%
_{3}$O$_{7}$ (a comprehensive review of the data published before 1992 can
be found in Ref.\cite{rad}, and many of more recent references are discussed
in Ref.\cite{PM}) in a decision-tree manner, eliminating the models
substantially incompatible with established experimental facts, eventually
focussing on the $pf\sigma $ hybridization models 
 and listing on a verbal level the interpretations of the existing
experiments as they emerge from such models.

The first three experimental facts that I want to emphasize are
\begin{quote}
1. Normal state conductivity drops sharply with the Pr concentration $x,$
and the material becomes essentially insulating at concentrations comparable
to the concentration at which superconductivity is fully suppressed.\\
2.  The dependence of $T_{c}$ on $x$ does not follow the classical
Abrikosov-Gor'kov law for superconductivity suppression by pair-breaking
impurities.\\
3.  In samples where $T_{c}$ is suppressed by alloying with Pr, it can be
restored back by partial substitution of Y by Ca, which adds holes to the
system.\\
\end{quote}
These three observations essentially eliminate the `pair breaking model'
where suppression of $T_{c}$ is ascribed to the pair-breaking effect of Pr,
acting like an impurity. The rational of this model is that Pr electrons are
to some extent present at the Fermi level, as witnessed by the unusually
large Neel temperature, and that for the $d-$wave superconductivity all
impurities are pair-breaking. Let me now explain
why the experimental facts above eliminate this possibility:

(i)  Theory of the pair-breaking impurity scattering is known in details.
Characteristic parameter is $\tau \Delta =2\pi \xi /l,$ where $\Delta $ is
the superconducting order parameter, $\xi $ is coherence length, $l=\tau
v_{F}$ is the mean free path for the pair breaking impurity scattering. On
the other hand, carrier localization due to impurities occurs when $%
l\sim a$ (lattice parameter), which is nearly two orders of magnitude stronger
condition on $l$ than the Abrikosov-Gor'kov condition $2\pi \xi /l\sim 1.$
Thus one expects superconductivity to be suppressed well before the normal
state conductivity becomes nonmetallic, as it happens in conventional
superconductors with magnetic impurities.

(ii)  The Abrikosov-Gor'kov law is very universal and particularly robust at
small impurity concentration. In particular, the law states that $T_{c}$
suppression is linear in impurity concentration. This is compatible with the
measurements on some $RE_{1-x}$Pr$_{x}$Ba$_{2}$Cu$_{3}$O$_{7}$ compounds (%
{\it e.g.}, Nd$_{1-x}$Pr$_{x}$Ba$_{2}$Cu$_{3}$O$_{7}),$ but not on the
others (like Y$_{1-x}$Pr$_{x}$Ba$_{2}$Cu$_{3}$O$_{7}),$ where suppression
starts essentially quadratically.

(iii)  Substituting Y with Ca, with the reaining composition unchanged,
 drives the material
into the so-called ``overdoped regime'' (hole concentration above the optimum),
which is characterized by a lower $T_{c}$  and a
smaller order parameter $%
\Delta $ than the ``optimally doped'' material, YBa$_{2}$Cu$_{3}$O$_{6.92}.$
If Pr suppresses the superconductivity via
pair-breaking, and does not change in the number and/or
character of the carriers, it should suppress $T_{c}$ further, not less,
in (Y,Ca)$_{1-x}$Pr$_{x}$Ba$_{2}$Cu$_{3}$O$_{7}$ compared with Y$_{1-x}$Pr$%
_{x}$Ba$_{2}$Cu$_{3}$O$_{7}.$

Having established that impurity pairbreaking is {\it not} the cause of the $%
T_{c}$ suppression, we are left with two options: (a) the pairing
interaction, whatever it may be, is weakened by the Pr doping, or (b) the
number or the character of the charge carriers changes. While I am not aware
of any experiment directly eliminating the first possibility, it is
disfavored by the fact (iii) above. Moreover, we know that Pr does not donate
electrons into the superconducting $pd\sigma ^{*}$ Cu-O band (``four-valent
Pr model''), because
\begin{quote}
4.  in the oxygen reduced samples, Y$_{1-x}$Pr$_{x}$Ba$_{2}$Cu$_{3}$O$%
_{7-\delta },$ $T_{c}$ is suppressed with $x$ slower than in fully
oxygenated $(\delta =7)$ samples.
\end{quote}
 The next illuminating experimental fact is 
\begin{quote}
5. unusually high N\'{e}el
temperature of PrBa$_{2}$Cu$_{3}$O$_{7}$, $T_N\approx 14$
 K (all other $RE$Ba$_{2}$Cu$_{3}
$O$_{7}$ have $T_{N}$ of a few K). This indicates that Pr $f$ states are
present at the Fermi level. This has direct connection with the $T_{c}$
suppression because\\
6. this suppression is enhanced when external pressure is applied (of
course, compression reduces the Pr-Cu and Pr-O bond lengths and increases
the corresponding hybridizations). This tells us that the relevant
interactions are either Pr($f)$-O($p)$ or Pr($f)$-Cu($d).$\\
\end{quote}

There are two messages in this findings: (a) The relevant hybridization of
the $f$ states is with some states close to Fermi level, because the Nd($f$)
states are farther away from $E_{F}$ than Pr($f$), so that larger hopping is
required for the same hybridization effect on the Cu-O states; however, (b)
the Pr-CuO hybridization should be stronger than some threshold for the
suppression to appear, thus the relevant hybridization is not with the
superconducting CuO $pd\sigma ^{*}$ band, but with some other states,
initially not at the Fermi level. The main question now is to identify those
states. It was done by Fehrenbacher and Rice\cite{FR}, who noted that in the
YBCO structure the nearest neighbors for a $RE$ atom are 8 oxygens forming
nearly an ideal cube, and that there is one particular $f$ orbital, $%
(x^{2}-y^{2})z$ , which has 8 lobes pointing exactly at those 8 oxygens,
allowing for $pf\sigma $ interaction. It was verified by LDA+U calculations
(which is a very good approximation for $f$ electrons) that indeed
\begin{quote}
7.  this orbital is the one (and the only one) occupied in PrBa$_{2}$Cu$%
_{3}$O$_{7}$ \cite{my}, and that the largest Pr-CuO hopping is in the $%
pf\sigma $ channel.
\end{quote}
It was furthermore suggested\cite{FR} and checked by calculations \cite{my}
that the total number of Cu-O holes changes little upon Pr doping, but
rather their character does; namely, that the $pd\sigma ^{*}$ holes are
being transferred to another state which we will call tentatively $pf\sigma
^{*}$ (because the transfer is caused by the $pf\sigma $ hopping), keeping
in mind however that admixture of other states, first of all of the $pd\pi
^{*}$ states, is not excluded by symmetry.

This chain of arguments, based on experimental data and numerical
calculations, essentially eliminates all other models except for the $p-f$
hybridization model. There are, however, open questions about the details of
this model. In particular, a key question is whether without
crystallographic disorder the ``new'' states are itinerant or localized? The
facts seem to point in the opposite directions:

\begin{itemize}
\item  Large indirect exchange suggest itinerancy. \underline{However}, it
might be that that the indirect exchange is not the only reason for high
Neel temperature.

\item  LDA+U calculations render a highly itinerant state. \underline{However%
}, LDA has a tendency to overestimate hoppings.

\item  PrBa$_{2}$Cu$_{3}$O$_{7}$ is stoichiometric, and it is disputed
whether the Pr-Ba disorder is sufficient to localize the carriers in the $%
pf\sigma ^{*}$ band. \underline{However}, the observation of
superconductivity in PrBa$_{2}$Cu$_{3}$O$_{7}$ suggests that it probably 
{\it is} a metal.
\end{itemize}

Fortunately, there is a nearly unambiguous clue in favor of the itinerant
model. Namely, there is a stunning observation which was at first received
as a total mystery, that the suppression rate of $T_{c}$ in different
members of the $RE_{1-x}$Pr$_{x}$Ba$_{2}$Cu$_{3}$O$_{7}$ family is entirely
different. For instance, $T_{c}$ is suppressed by less than 40\% in Y$_{0.7}$%
Pr$_{0.3}$Ba$_{2}$Cu$_{3}$O$_{7}$ or Yb$_{0.7}$Pr$_{0.3}$Ba$_{2}$Cu$_{3}$O$%
_{7},$ while in Nd$_{0.7}$Pr$_{0.3}$Ba$_{2}$Cu$_{3}$O$_{7}$ the suppression
is nearly complete. Of course, without Pr doping all these compounds have
nearly the same $T_{c}.$ Of course, a state localized at a given Pr ion does
not know about what $RE$ is sitting in the other cells, so one concludes
that the ``hole-grabbing'' state which appears due to the $pf\sigma $
hybridization should be itinerant. A question of secondary importance
remains of whether this state exists in the undoped $RE$Ba$_{2}$Cu$_{3}$O$%
_{7}$, but is fully occupied, and the role of the  $pf\sigma $ hybridization
is merely to move this state closer to the Fermi level, or the  $pf\sigma $
hybridization {\it creates} this state. The first case corresponds to the
LDA+U calculations of Ref.\cite{my}, where the state in question is a $pd\pi
^{*}$ band, with the width of about 1/2 of that of the $pd\sigma ^{*}$ band,
and lies entirely under the Fermi level in all $RE$Ba$_{2}$Cu$_{3}$O$_{7}$
except for $RE=$Pr. The neglect of the $pd\pi $ hopping leads to the second
case, when the width of the  ``hole-grabbing'' band is proportional to the $%
pf\sigma $ hopping\cite{note}. The linear $T_{c}$ suppression in Nd$_{1-x}$Pr%
$_{x}$Ba$_{2}$Cu$_{3}$O$_{7}$ is easier to explain in the first case, but
the second gives better agreement\cite{my} with the measured number of the O(%
$p_{z})$ holes\cite{merz}. For most experiments the difference between the
two cases is insignificant, and the conclusions can be formulated as follows:

\begin{itemize}
\item  With Pr doping, electrons are transferred from the usual
superconducting  $pd\sigma ^{*}$ band to a new band, which includes $pd\pi
^{*}$  and/or $pf\sigma ^{*}$ states. 

\item  The new band is (a) heavy and (b) likely to be strongly renormalized
by magnetic interactions.

\item  Because of that, the carriers in this band are easily localized.

\item  As a consequence, we expect this band to show its metallic character
only in well-ordered, stoichiometric samples, and probably only at a short
length scale.
\end{itemize}

Regarding the newly observed superconductivity\cite{newT} in
PrBa$_{2}$Cu$_{3}$O$_{7},$ it appears to be a more novel superconductor than
all other cuprate high $T_{c}$ materials known: it is the only one where
superconducting carriers are not residing in the Cu$(x^{2}-y^{2})-$O$%
(p_{\sigma })$ bands, but are of entirely different character. One can ask
why are the critical temperatures in the two compounds, YBCO and PBCO, so
similar? In the framework of the suggested model it is a sheer coincidence,
which should be removed, for instance, by external pressure. Indeed, it was
observed recently\cite{P} that the pressure coefficient $dT_{c}/dP$ in PBCO
is an order of magnitude larger than in any other $RE$Ba$_{2}$Cu$_{3}$O$_{7}$%
. Interestingly, in the $df\sigma $ hybridization model one expects $%
dT_{c}/dP$ to be negative in $RE_{1-x}$Pr$_{x}$Ba$_{2}$Cu$_{3}$O$_{7}$ at
small $x$, because pressure increases hybridization and thus the charge
transfer from the $pd\sigma^* $ to the $pf\sigma^*-pd\pi^*
$ states, while the same
argument predicts $dT_{c}/dP$ to be positive in PrBa$_{2}$Cu$_{3}$O$_{7}$
(since in this system the superconductivity occurs in the $pf\sigma^*
-pd\pi^* $ band
itself). Both predictions are in agreement with the experiment. Finally, I
would like to point out recent measurements\cite{Com} that yielded
qualitatively different Compton profiles in YBa$_{2}$Cu$_{3}$O$_{7}$ and PrBa%
$_{2}$Cu$_{3}$O$_{7}$. This usually indicates different topology of the
Fermi surface and is in agreement with the concept of qualitatively
different carriers in this two compounds.

This work was supported by the Office of Naval Research. I am thankful
to V.N. Narozhnyi for his useful comments.
\vskip .2cm
\hrule
\vskip .2cm

{\bf Appendix.}
The following cartoons illustrate how the $pf\sigma-pd\pi$ hybridization
model explains the main experimental facts.\\
\centerline{\epsfig{file=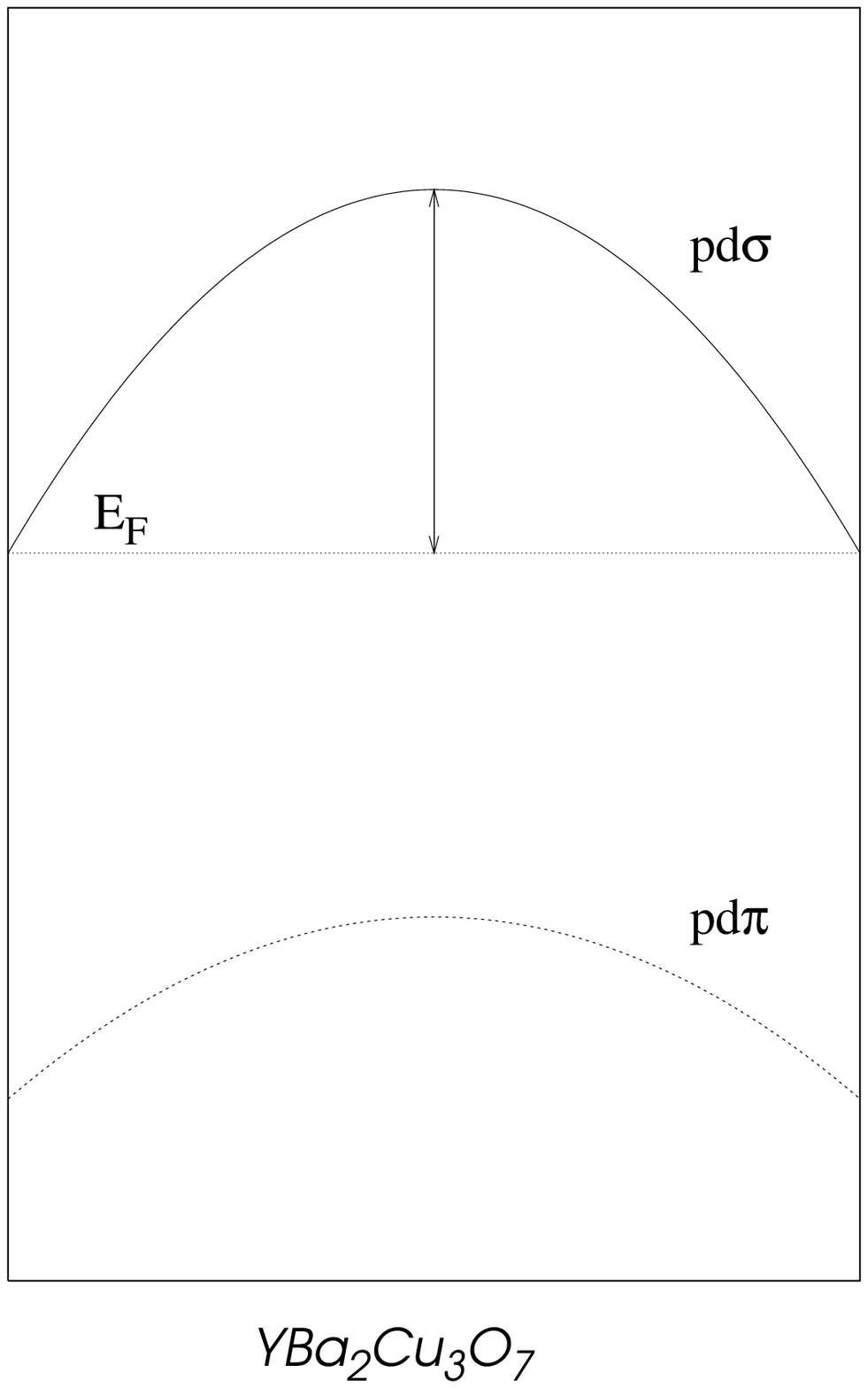,width=0.25\linewidth}
\epsfig{file=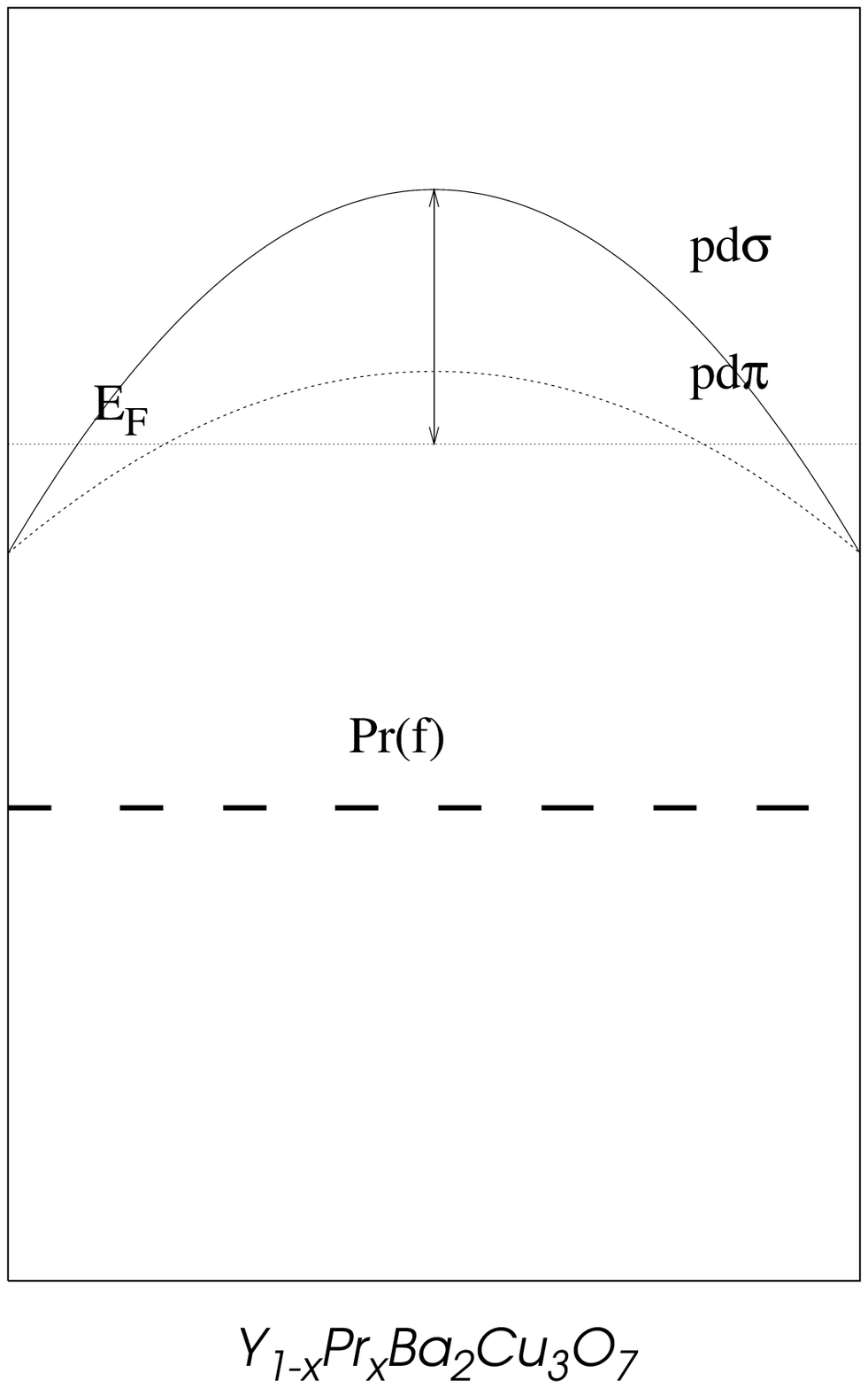,width=0.25\linewidth}
\epsfig{file=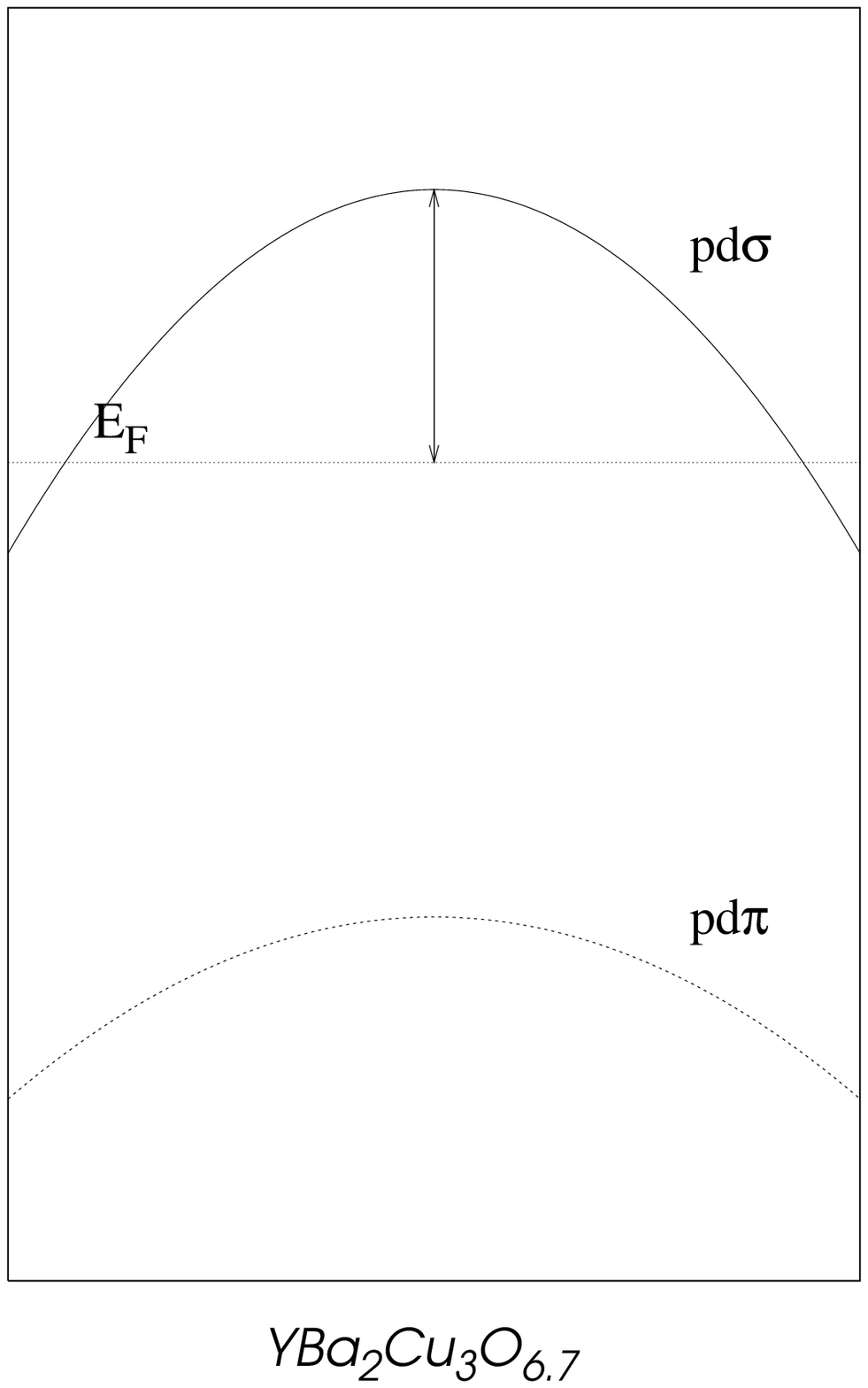,width=0.25\linewidth}
\epsfig{file=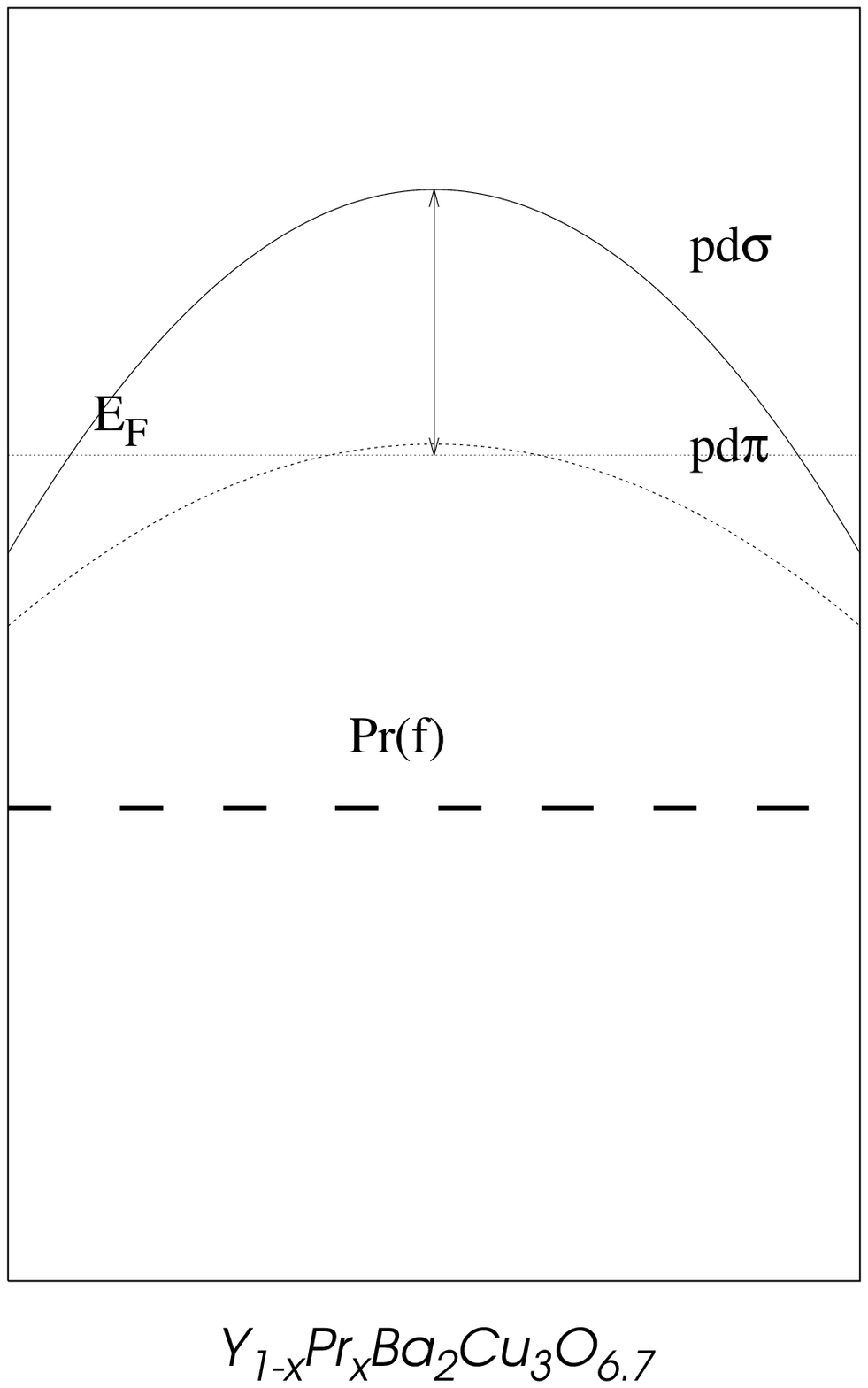,width=0.25\linewidth} }

(a) In  YBa$_2$Cu$_3$O$_7$ the $pd\pi^*$ band is fully
occupied and does not influence transport properties or superconductivity.
(b) In Y$_{0.8}$Pr$_{0.2}$Ba$_2$Cu$_3$O$_7$ this band is pushed up
and grabs some holes from the superconducting $pd\sigma^*$ band.
(c,d) Since in YBa$_2$Cu$_3$O$_{6.7}$ the Fermi level
is higher than in YBa$_2$Cu$_3$O$_7$, the same shift of the
$pd\pi^*$ band produces smaller changes in the hole
concentration in the  $pd\sigma^*$ band.
\vskip 1cm
\centerline{\epsfig{file=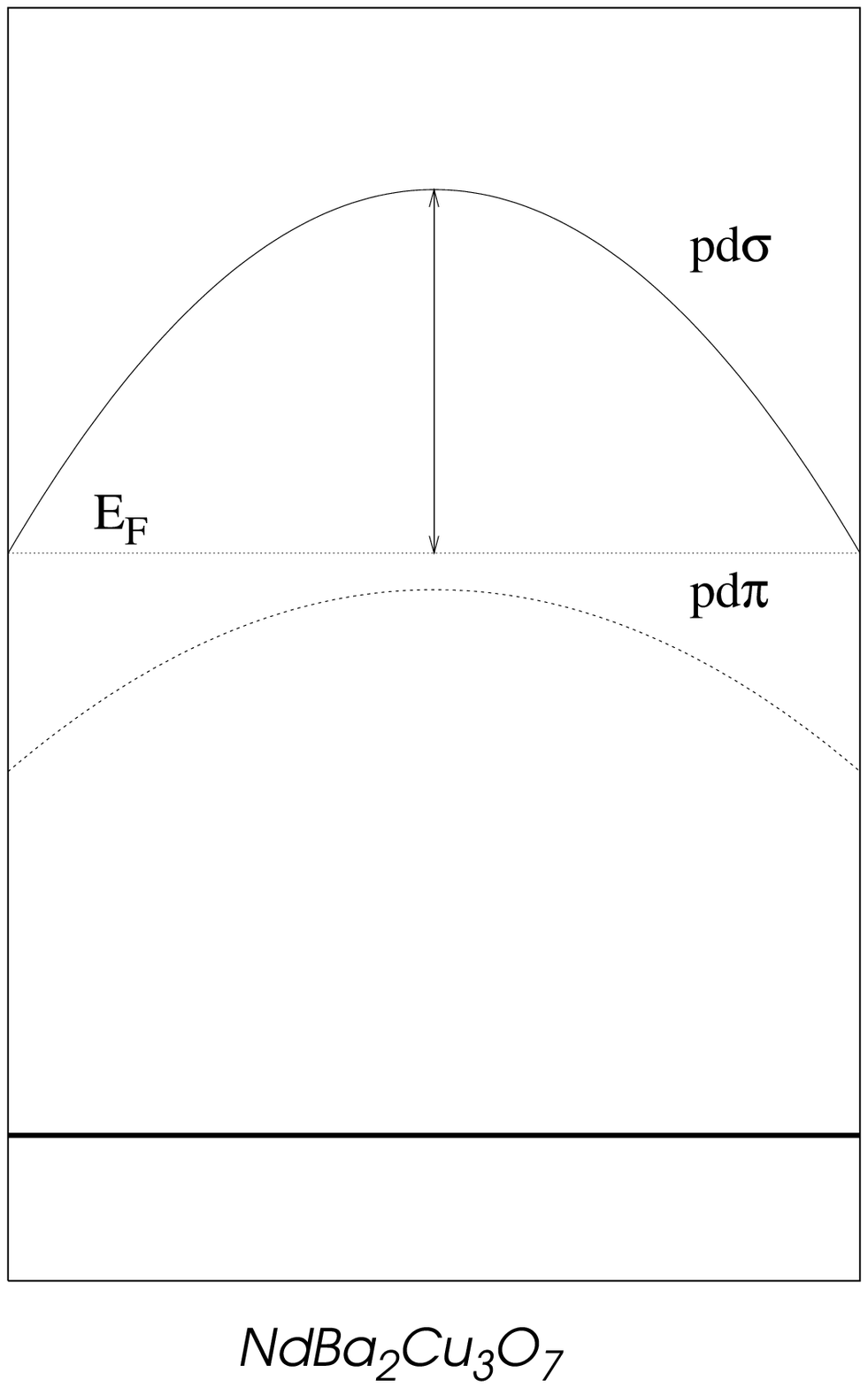,width=0.25\linewidth}
\epsfig{file=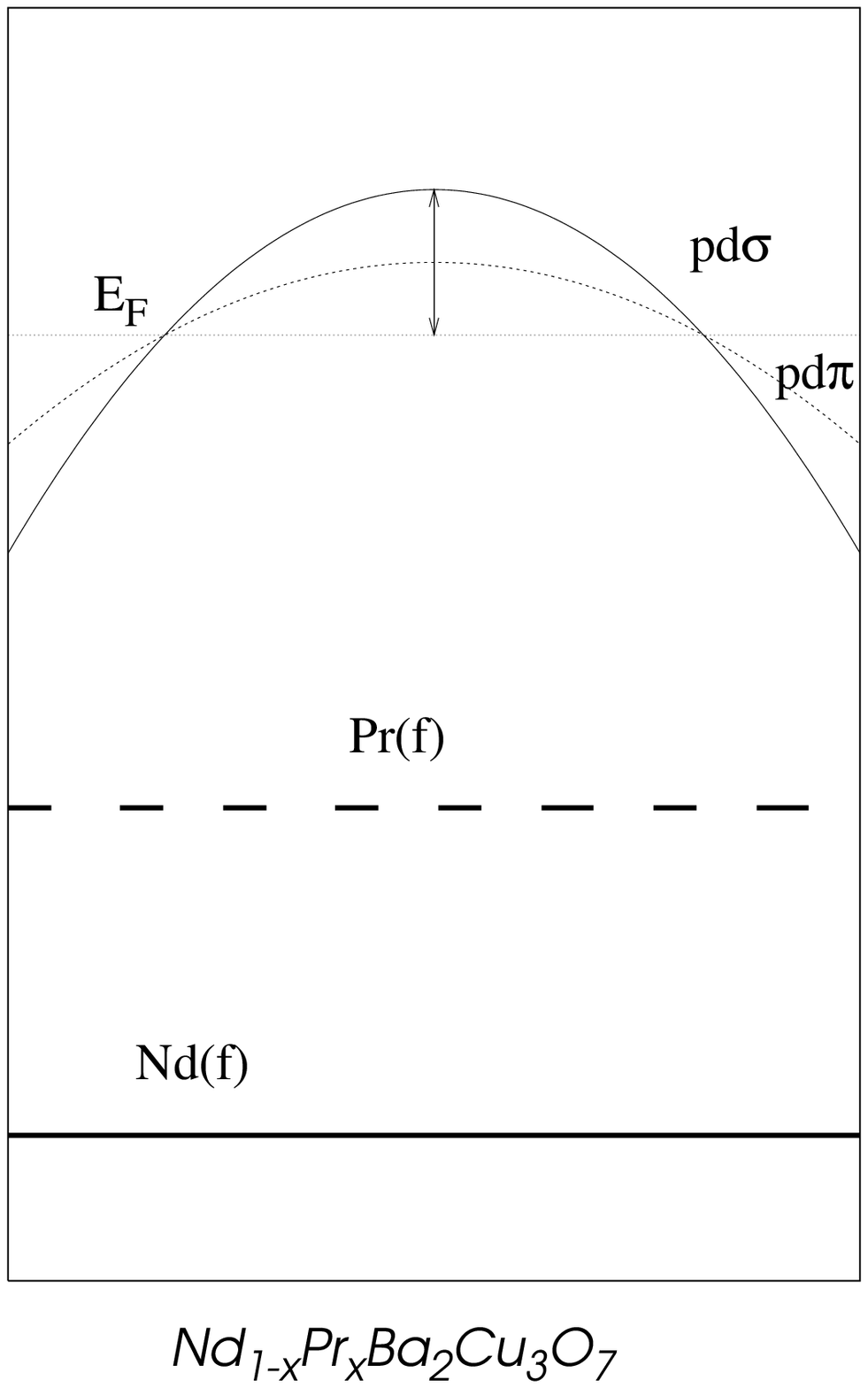,width=0.25\linewidth}
\epsfig{file=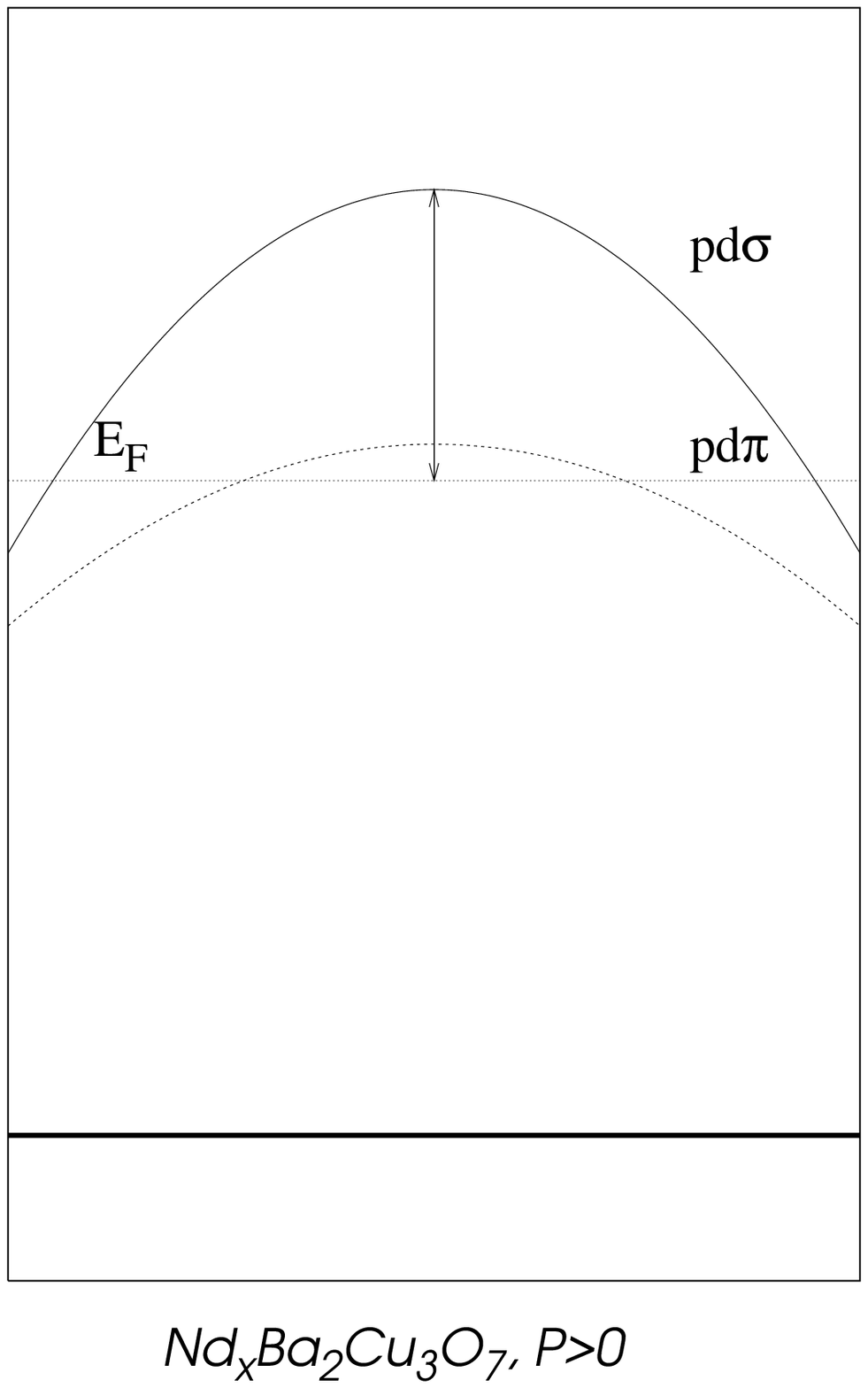,width=0.25\linewidth}
\epsfig{file=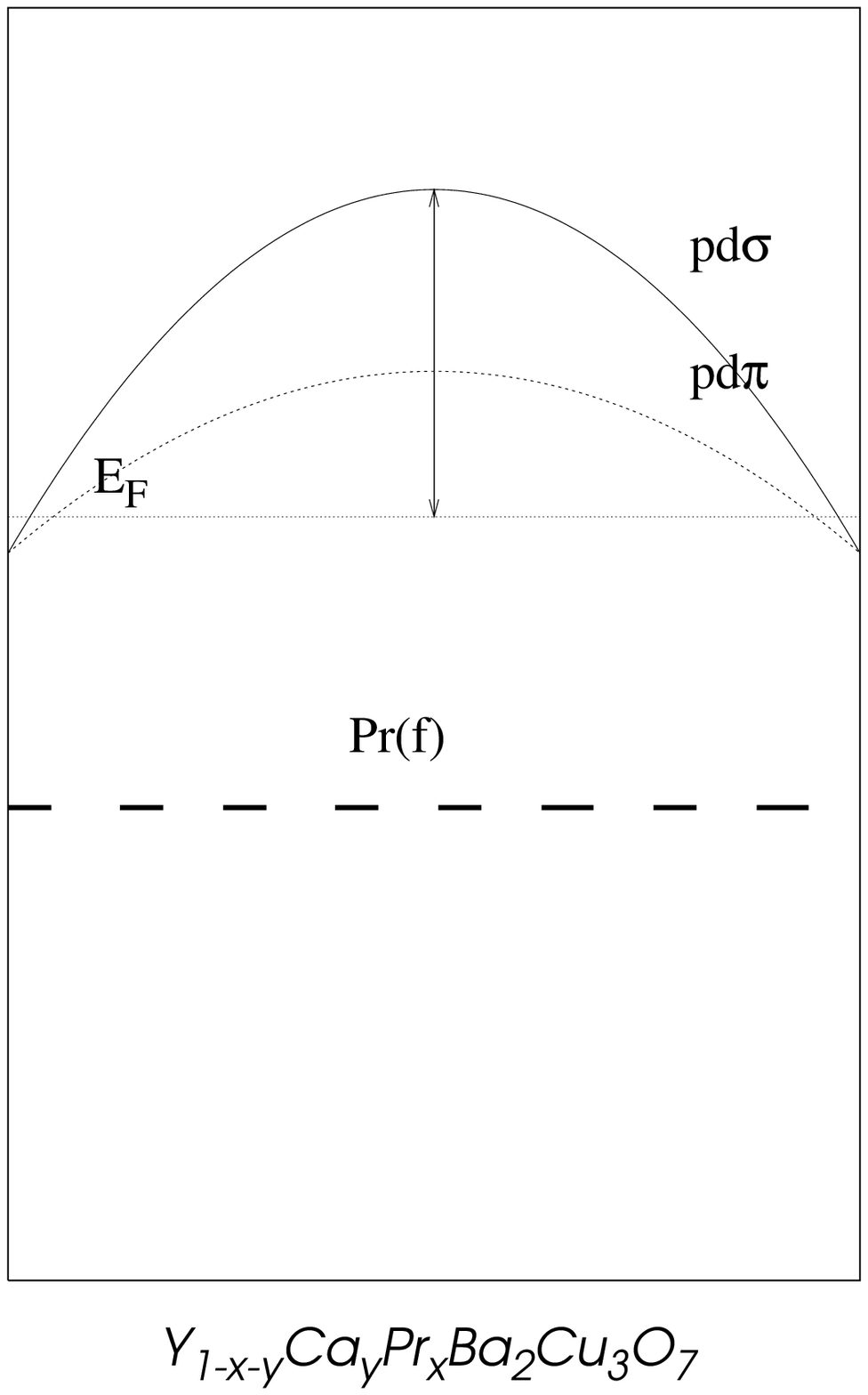,width=0.25\linewidth} }

(e) In NdBa$_2$Cu$_3$O$_7$ the $pd\pi^*$ is pushed up by the $pf\sigma$
interaction, but barely not enough to cross the Fermi level.
(f) In Nd$_{0.8}$Pr$_{0.2}$Ba$_2$Cu$_3$O$_7$ this band
lies higher than in Y$_{0.8}$Pr$_{0.2}$Ba$_2$Cu$_3$O$_7$, because it was
substantially higher even without Pr; furthermore,
(g) external pressure increases the $pf\sigma$ hybridization;
At a sufficently high pressure
 even in the pure NdBa$_2$Cu$_3$O$_7$ the $pd\pi^*$
band crosses the Fermi level, and a suppression of $T_c$
is expected.
(h) Doping with Ca introduces additional holes into the system
which offset the depleting effect of the  $pd\pi^*$ band on
the $pd\sigma^*$ band.
\vspace*{\fill}
\eject
These cartoons illustrate how external pressure increases the $T_c$
in PrBa$_2$Cu$_3$O$_7$.
\vskip 1cm
\epsfig{file=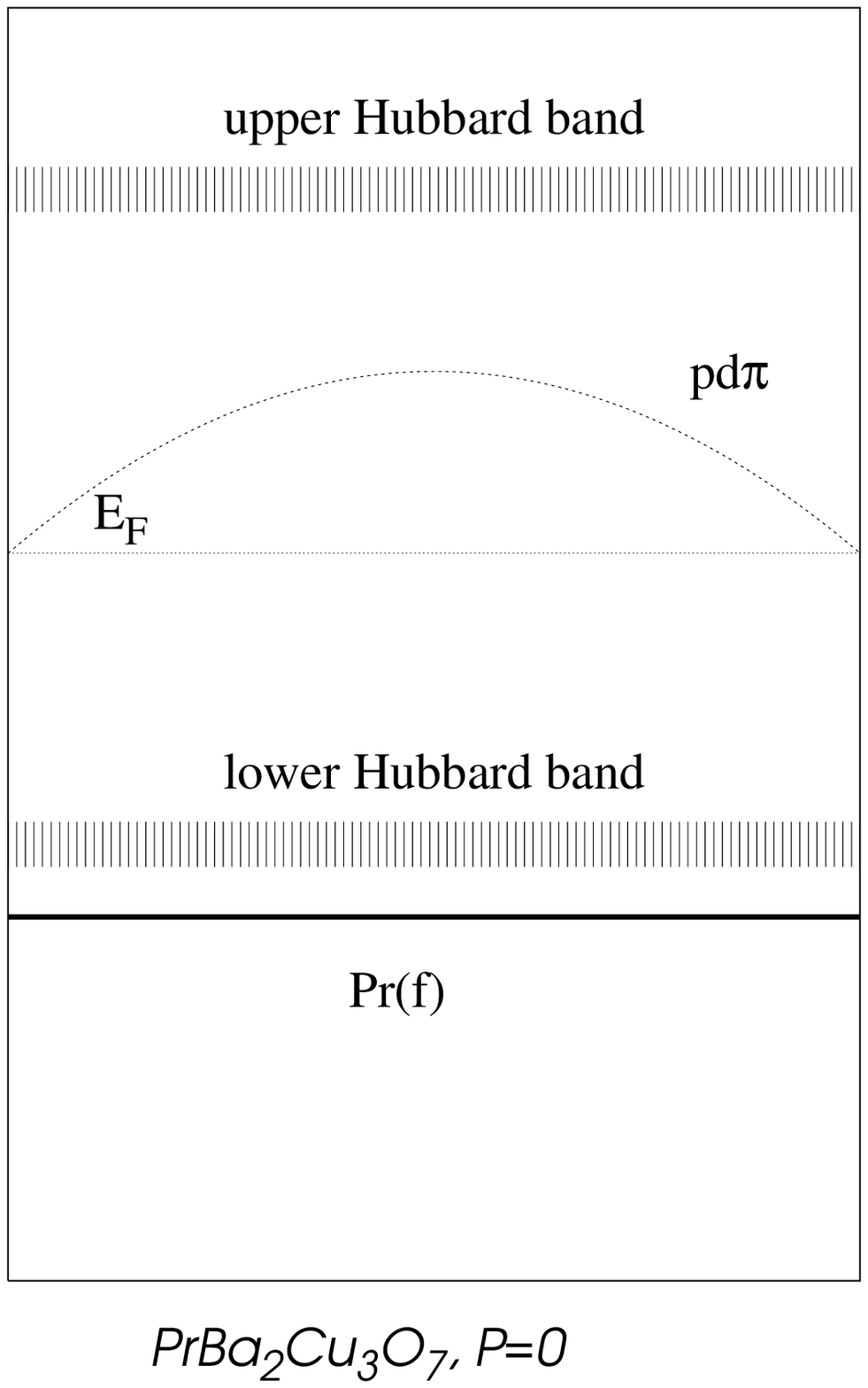,width=0.25\linewidth}
\epsfig{file=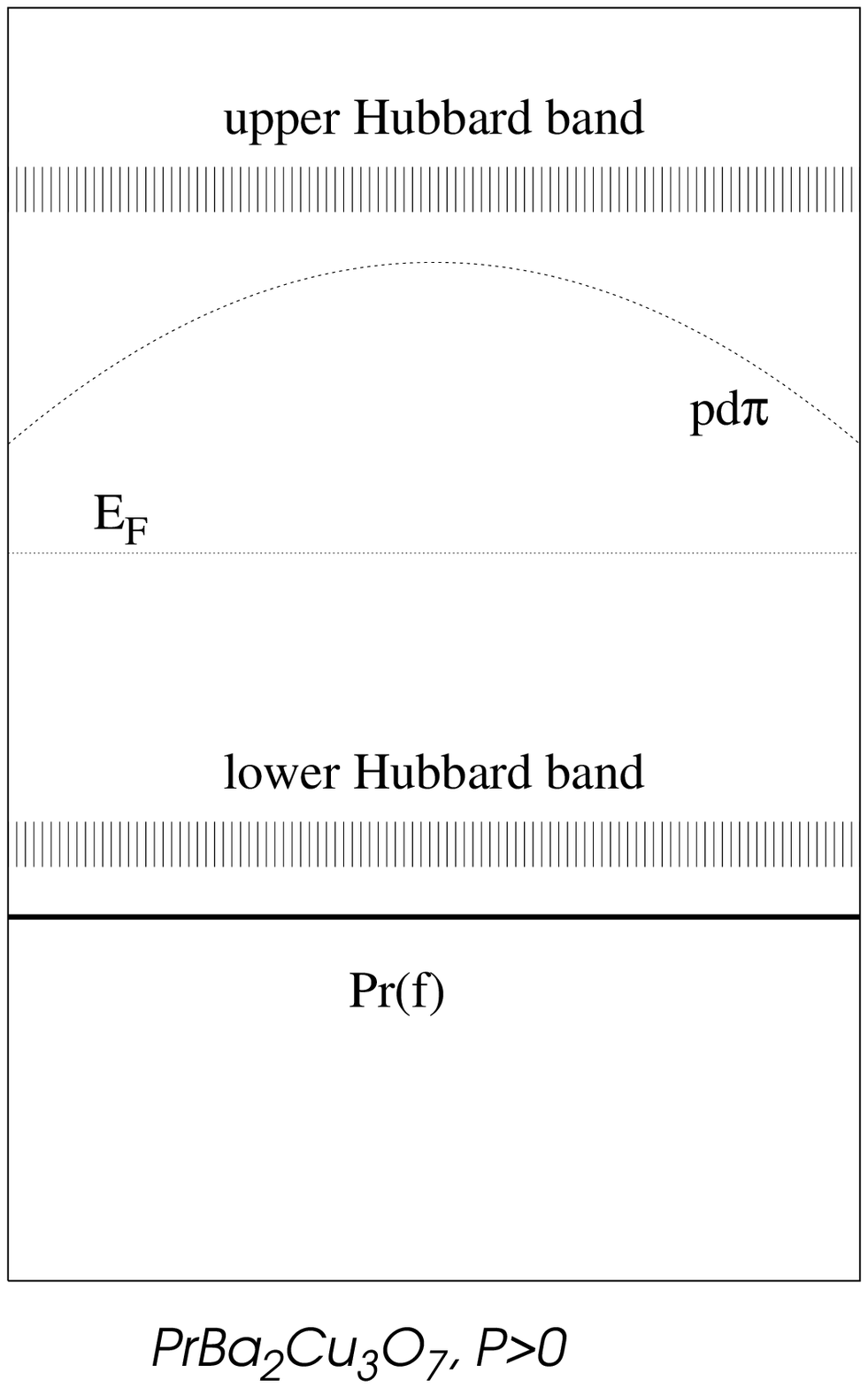,width=0.25\linewidth}
\begin{minipage}{0.4\linewidth}
\vskip -5cm
(i) When Y is fully substituted by Pr, the $pd\sigma^*$ band
is so close to half-filling that it undergoes the Mott-Hubbard
transition. The conductivity and superconductivity is carried 
instead by the holes in the new $pd\pi^*$ band.
(j) At  a higher pressure, due to increased $pf\sigma$
interaction, the position of this band is higher than at $P=0$,
and thus the number of holes in this band is larger.
\end{minipage}

\end{document}